\documentclass{article}

\usepackage{microtype}
\usepackage{graphicx}
\usepackage{subfigure}
\usepackage{booktabs} 

\usepackage[preprint]{neurips_2025}

\usepackage{xurl}

\usepackage{hyperref}

\usepackage{amsmath}
\usepackage{amssymb}
\usepackage{mathtools}
\usepackage{amsthm}

\usepackage{paralist}

\usepackage[capitalize,noabbrev]{cleveref}

\theoremstyle{plain}

\theoremstyle{definition}

\theoremstyle{remark}

\usepackage[textsize=tiny]{todonotes}

\usepackage{xspace}
\def\systemname#1{\textsf{#1}\xspace}
\newcommand{\lc}{\systemname{leanCoP}}
\newcommand{\rlc}{\systemname{rlCoP}}
\newcommand{\plc}{\systemname{plCoP}}
\newcommand{\gc}{\systemname{graphCoP}}

\usepackage{pgfplots}
\pgfplotsset{compat=1.17}

\newcommand{\dataset}{D}
\newcommand{\myinput}{\boldsymbol x}
\newcommand{\mylabel}{\boldsymbol y}
\newcommand{\myoutput}{d}
\newcommand{\outdim}{m}
\newcommand{\numallowed}{k}
\newcommand{\distribution}{\mathcal{P}}

\newcommand{\prploss}{\mathsf{Libra}\text{-}\mathrm{loss}}
\newcommand{\prplosssubscript}{\mathrm{Lib}}

\newcommand{\nllloss}{\mathsf{NLL}\text{-}\mathrm{loss}}

\newcommand{\uloss}{\mathrm{Uniform}\text{-}\mathrm{loss}}

\newcommand{\bmeritloss}{\mathrm{\beta}\text{-}\mathrm{merit}\text{-}\mathrm{loss}}

\newcommand{\bmeritlossB}{\mathrm{0.5}\text{-}\mathrm{merit}\text{-}\mathrm{loss}}

\newcommand{\bmerit}{\mathrm{\beta}\mathrm{M}}

\newcommand{\x}{\mathbb{X}} 

\newcommand{\nll}{\mathrm{NLL}}

\newcommand{\nsamples}{n}

\newcommand{\mylabelscalar}{y}
\newcommand{\truetarget}{\boldsymbol y_{\textrm{true}}}

\newcommand{\seqlen}{\ell}

\newcommand{\myinputvar}{\myinput}
\newcommand{\mylabelvar}{\mylabel}
\newcommand{\truetargetvar}{\truetarget}

\newcommand{\outputspace}{\mathcal{D}}
\newcommand{\Loss}{\mathcal{L}}

\newcommand{\probs}{{\boldsymbol p}}
\newcommand{\probsscalar}{{p}}					
\newcommand{\classifier}{\boldsymbol f}			

\newcommand{\pacc}{P_{\textrm{acc}}}

\title{Partial Label Learning for Automated Theorem Proving}

\author{%
  Zsolt Zombori \\
  HUN-REN Alfr\'{e}d R\'{e}nyi Institute of Mathematics\\
  E{\"o}tv{\"o}s Lor\'{a}nd University\\
  Budapest, Hungary \\
  \texttt{zombori@renyi.hu} \\
  \And
  Bal\'{a}zs Indruck\\
  HUN-REN Alfr\'{e}d R\'{e}nyi Institute of Mathematics\\
  Budapest, Hungary \\
  \texttt{bbbindruck@gmail.com} \\
}

\begin{document}

\maketitle


\begin{abstract}
We formulate learning guided Automated Theorem Proving as Partial
Label Learning, building the first bridge across these fields of
research and providing a theoretical framework for dealing with
alternative proofs during learning. We use the \plc theorem prover to
demonstrate that methods from the Partial Label Learning
literature tend to increase the performance of learning assisted
theorem provers.
\end{abstract}

\section{Introduction}
\label{sec:introduction}
We formulate learning guided Automated Theorem Proving as
Partial Label Learning, building the first bridge across these
two fields of research, providing novel tools for the former and an
important use case for the latter.

Partial Label Learning~\citep{pllsurvey} deals with learning in the
presence of imperfect supervision, where in each training sample the
input is associated with a set of labels, one of which is the true
label. This noisy supervision could originate, for example, from labelling
errors, from the presence of hidden variables or from synthetic label
generation.

We focus on learning to guide theorem proving systems where the
goal is to find a sequence of atomic inferences that constitute a
valid proof of some target problem in the expentially growing space of
possible derivations. An often overlooked aspect of this training task
is that theorems have many proofs, several of which could be
available during training time. Our paper aims to directly address
this aspect of the learning task, trying to identify the proper way of
exploiting alternative proofs, as well as failed derivations.

Given a dataset of target theorems, each of which is associated with a
set of known proofs, training a proof search guidance model can be
directly interpreted as a Partial Label Learning task with some
unusual properties. First, proofs are typically represented as
sequential, structured objects, as opposed to atomic
predictions. Second, the set of possible derivations is infinite,
while Partial Label Learning typically assumes a small and fixed set
of possible predictions.
Nevertheless, as we shall show, these differences can be overcome.

Previous work has introduced heuristic ways of dealing with more
proofs, often resorting to using the shortest proof or a random
selection. Here, we attempt to provide a systematic study of how best
to exploit the existence of alternative proofs, examining several learning
methods that have recently been proposed in the PLL literature.

We use the \plc theorem prover to demonstrate the usefulness of some
methods from Partial Label Learning. However, these methods are not
system specific and are easily available for several provers that
rely on inference level guidance. To the best of our knowledge, this
is the first work focused on alternative proofs for internal guidance.
We expect no expertise from the reader on the topic of Automated
Theorem Proving. Abstracting away from many of its subtleties, we
characterise the problem as a Reinforcement Learning environment and
focus on the learning aspect.
We provide several experiments with the \plc system. Our results seem
to corroborate previous findings that when training on a single proof,
the shortest proof tends to work best. Furthermore, we find that the
Partial Label Learning methods that learn jointly from all the proofs
and aim to strike some balance among them, consistenly yield further
improvement over previous approaches.

The rest of the paper is structured as
follows. Section~\ref{sec:prelims} gives an overview of relevant
concepts and related works. In Section~\ref{sec:pll_guidance} we
formulate learning to guide proof search as a Partial Label Learning
task. In Section~\ref{sec:pll_plc} we introduce the \plc system
and describe how its training task fits into
the Partial Label Learning framework. Section~\ref{sec:pll_methods}
introduces some prominent methods for Partial Label
Learning. Section~\ref{sec:experiments} presents our experimental
results. Finally we conclude the paper in
Section~\ref{sec:conclusion}.

\section{Preliminaries and Related Work}
\label{sec:prelims}

We briefly introduce Partial Label Learning and internal guidance based
theorem proving.

\subsection{Partial Label Learning}

\emph{Partial Label Learning (PLL)} is related to supervised
classification, but it assumes weaker supervision.  Supervised
classification is the task of learning a function that conforms to a
given set of samples $\dataset =\big\{\big(\myinput^{(j)},
\mylabel^{(j)}\big)\big\}_{j=1}^{\nsamples}$ where $\myinput \in \x$
is the input and $\mylabel \in \{0,1\}^\outdim$ is the one-hot encoded
desired output.
In PLL, however, there
can be more than one allowed output, represented as $\mylabel$ having
multiple entries being 1.  PLL assumes one single correct output among
the given 1 entries that is unknown at training time (thus the labels
are uncertain).  We use $\numallowed$ to denote the number of
acceptable outputs associated with label $\mylabel$ in supervision.
We use $\truetarget$ to denote the one-hot encoded unknown correct
output. Note that the assumption of a single correct output only plays
a role in theoretical investigations: it does not influence learning
as it is unknown during training.

PLL assumes a joint generating distribution $\distribution(\myinputvar, \truetargetvar,
\mylabelvar)$ 
on inputs $\myinputvar \in \x$, true one-hot outputs $\truetargetvar \in \{0,1\}^\outdim$,
and partial supervision $\mylabelvar \in \{0,1\}^\outdim$. 
In other words, the observed labels $\mylabelvar$ are a distorted representation of 
the true labels $\truetargetvar$.
The goal is to learn a
function $\classifier$ in a given target class 
that maximizes

$\mathbb{E}_{\distribution(\myinputvar,\truetargetvar,\mylabelvar)}[P(\classifier(\myinput)=\truetarget)]$. 

We focus on
classifiers that output a probability distribution $\probs$ over
output space $\outputspace$.
PLL is very general and a number of well-studied problems can be seen
as particular instances.  Over the last decade a multitude of
proposals for PLL have emerged, see for example \citet{pllsurvey} for
a recent survey.  One prominent approach, referred to as
``average-based'' involves methods that average the model's
performance over the allowed labels \citep{pllavgbased,pllleveraging}.
Other methods, called ``identification-based'', attempt to learn the
noise model in combination with the prediction
\citep{pllzoubin,pllearly,pllidentificationbased,pllanotheridentificationbased,provablyconsistentpll}.

In many cases, the differences among methods boil down to using
different loss functions for optimisation.
These
functions are often derived using specialised assumptions about the
data generating process. A different approach is taken in
\citet{libra}, which makes no assumptions about data generation and
analyses several loss functions from the point of view of
learning dynamics. Our paper explores how changes in the loss function
influences learning in the context of Automated Theorem Proving.

\subsection{Internal Guidance for Theorem Proving}
In \emph{Automated Theorem Proving (ATP)}, we are given a set of
axioms along with a conjecture and our goal is to find a proof
certifying that the conjecture must hold whenever the axioms
hold. Most ATP systems work by formulating a large search space of
potential derivations and then proceed by some sort of systematic
search. However, the size of the search space poses a fundamental
challenge. Decades of research has lead to the accumulation of
heuristics that help to reduce the search space and there is also
growing interest in using Machine Learning (ML) to acquire such
heuristics directly from theorem proving data.  Many components of the
proof search enterprise can be enhanced via learning, e.g.,
\emph{premise selection}~\citep{Urban07,deepmath,formulanet},
\emph{strategy
selection}~\citep{blistr,SchaferS15,KuhlweinU15,HoldenK21}, next
inference
selection~\citep{deep_guidance,enigma,graph_holist,lazycop,CrouseAMWCKSTWF21,plcop,flop,plcop_gnn}.
The latter task is called \emph{internal guidance} and our paper deals
with this approach.

In internal guidance, models are trained to evaluate a given state ---
called \emph{value} model --- or to evaluate valid inferences in a
given state --- called \emph{policy} model. The models can be trained
with supervision from human generated proofs or from the prover's own
proof attempts in a reinforcement learning fashion. Crucially, the
models reduce the effective branching factor, allowing the prover to
focus only on the more promising derivations.

This learning problem can be phrased as a Reinforcement Learning (RL)
environment. In this environment states are prover states and actions
are valid inferences. What makes theorem proving an extremely hard
environment is its reward scheme: there are no principled ways of
providing partial reward before a proof is found and it is poorly
understood how one proof can be considered more valuable than
another. Consequently, it is usually represented with sparse binary
rewards: the only time the agent receives a reward of $1$ is when it
finally finds the proof.

We investigate offline training a policy model $\classifier$ from a
database of successful and failed derivations. Given prover state
$\myinput$, the policy model produces a probability distribution
$\probs= \classifier(\myinput)$ over inferences applicable in state
$\myinput$. This task, as we shall show, can be interpreted as a PLL
task.

While our framework is applicable to many provers \citep[e.g.][]{enigma,
  deep_guidance}, we conduct our experiments using the latest version
of the \plc system \citep{plcop,plcop_gnn}, which is based on the \lc connection
calculus. \lc~\citep{leancop} is a compact theorem prover for
first-order logic, implementing connection tableau search.
A series of learning systems guiding the \lc connection
calculus have been developed. Of these, we highlight three that use
roughly the same reinforcement learning setup: \rlc~\citep{rlcop},
\plc~\citep{plcop,plcop_gnn} and
\gc~\citep{prop_invariant_embedding}. These systems search for proofs
using Monte Carlo Tree Search~\citep{mcts} and they train value and
policy functions similarly to
AlphaZero~\citep{alphazero,thinking_fast_and_slow}.

\subsection{Learning in the Presence of Alternative Proofs}

PLL for theorem proving can be seen as the study of learning in the
presence of alternative proofs. Often, learning based provers train on
all proofs found, without paying attention to which proofs are for the
same problem \citep[e.g.][]{rlcop,plcop}.  Possibly the first paper
that tackles the question of alternative proofs is
\citet{PAAR-2012:Learning_from_Multiple_Proofs}, in the context of
selecting relevant premises for the prover. The authors find that the
best results are obtained when learning from the single shortest
proofs --- even better than using human provided proofs --- and this
approach has become a bit of a folklore. \citet{atpboost} presents
another system trained for premise selection which also finds the
shortest proofs to be the most beneficial. In contrast with this,
\citet{flop} identifies several scenarios where the shortest proof is
too specific to the target problem and does not help generalising to
related problems. Our paper aims to shed more light on how best to
handle alternative proofs.

\section{PLL for Internal Guidance}
\label{sec:pll_guidance}

Given a theorem proving system, consider the set $\outputspace$ of all
finite derivation sequences. For any problem $\myinput$, any
derivation $\myoutput \in \outputspace$ can be grouped into one of
three categories: 1) \textbf{Proof}: $\myoutput$ is a proof of $\myinput$, 2) \textbf{Failure}: $\myoutput$ is a failed proof attempt, i.e., it doesn't prove $\myinput$ and it cannot be extended, 3) \textbf{Unknown}: $\myoutput$ is not a proof of $\myinput$, but there is at least one valid inference step extending it.


Given some canonical ordering of derivations $\myoutput_1, \myoutput_2
\dots \myoutput_i \dots$, each problem $\myinput$ can be associated with an
indicator vector $\mylabel \in \{0,1\}^{|\outputspace|}$, which is the
one-hot encoded set of proofs of $\myinput$, i.e., $\mylabel_i=1$ exactly
when the $i$th derivation $\myoutput_i$ is a proof of $\myinput$.

Consider $\dataset =\big\{\big(\myinput^{(j)},
\mylabel^{(j)}\big)\big\}_{j=1}^{\nsamples}$ of $\nsamples$ samples
where $\mylabel^{(j)}$ decodes the set of proofs of
$\myinput^{(j)}$. If we further assume that there is a single true
proof $\truetarget^{(j)}$ of $\myinput^{(j)}$ then we obtain a PLL
classification problem, in which the aim is to predict the true proof
of the input problem.
There are, however, several issues that make this task an unorthodox
PLL problem. We discuss each of them in the remaining of this section.

\paragraph{Sequential Prediction}
Instead of predicting distributions over entire derivations, we are
interested in policy models predicting distributions over atomic
inference steps. However, the policy model implicitly induces a
distribution over sequences as well.  Consider derivation $\myoutput =
i_1, i_2, \dots i_{\seqlen}$ with corresponding (preceding) prover states
$\myinput_1, \myinput_2, \dots \myinput_{\seqlen}$. Then, the
predicted probability of $\myoutput$ is $\prod_{j=1}^{\seqlen}
\classifier(\myinput_j)_{I(i_j)}$, where $I(i_j)$ refers to the index
in the output of $\classifier$ that corresponds to inference $i_j$.
Consequently, any PLL method that optimises the probabilities of
the derivations can be directly applied to the policy model which
produces the derivation probabilities in a sequential fashion.

\paragraph{Infinite Output Space}
Even if the set of possible atomic inferences is finite, the set of
derivations need not be. And even if we restrict our attention to
derivations up to a fixed length, the number of possible derivations
might still be too large to handle explicitly. We cannot enumerate all
derivations, not even all proofs. Consequently, we have to resort to
some sort of sampling, as done in \citet{guu-etal} for semantic
parsing. In practice, theorem proving datasets already provide a
sample since proof search discovers only a limited number of
derivations for a single problem.

\paragraph{No Single True Proof}
PLL assumes a single hidden, unknown true label that the model aims
to discover. In the case of theorem proving, it is not clear that
the single true proof is well defined, as all proofs are
equivalent in terms of certifying the validity of the target
problem. However, $\truetarget$ does not influence training (since
it is not known during training). It may be useful for evaluation,
but for theorem proving, we will evaluate models based on their
ability to help the prover find proofs --- instead of their ability
to predict $\truetarget$, so the exact definition of $\truetarget$
will not matter.


\section{PLL in the \plc system}
\label{sec:pll_plc}

\plc implements a learning method called \emph{Expert
Iteration}~\citep{expert_iteration} in the Reinforcement Learning
literature and often referred to as the \emph{MaLARea-style} learning
loop \citep{Urban07} in the theorem proving community. Learning
emerges as the interaction of two components: an \emph{Expert} system,
which searches for proofs and produces a tree of the explored portion
of the search space, and an \emph{Apprentice} system which learns to
imitate the Expert. The Expert's search is guided by Apprentice's
prediction and the output of the search is used to update the
Apprentice.

In \plc, the Expert is implemented as guided \emph{Monte Carlo Tree
Search (MCTS)}~\citep{mcts} and the Apprentice consists of a value and
a policy model.  MCTS produces a search tree in which each node ---
prover state --- is labelled with a value estimate (the average reward
accumulated in the subtree below the given node), and each edge ---
inference step --- is labelled with a probability estimate (the
frequency with which MCTS selected the given inference from the state
associated with its parent node). These estimates are used to train a
value and a policy model, based on \citet{alphazero}, which are used
to guide MCTS in the next iteration of proof search. Experiments
suggest \citep[see][]{rlcop} that the value model has minor effect on
the model's performance, most of the learning associated benefit comes
from the policy model. Hence, we focus only on policy
training.

The question we raise is how to best train the policy from the
explored search tree. We claim that learning to imitate MCTS
statistics is just one of the possible approaches.

\subsection{A Novel Look at ATP Training Data}

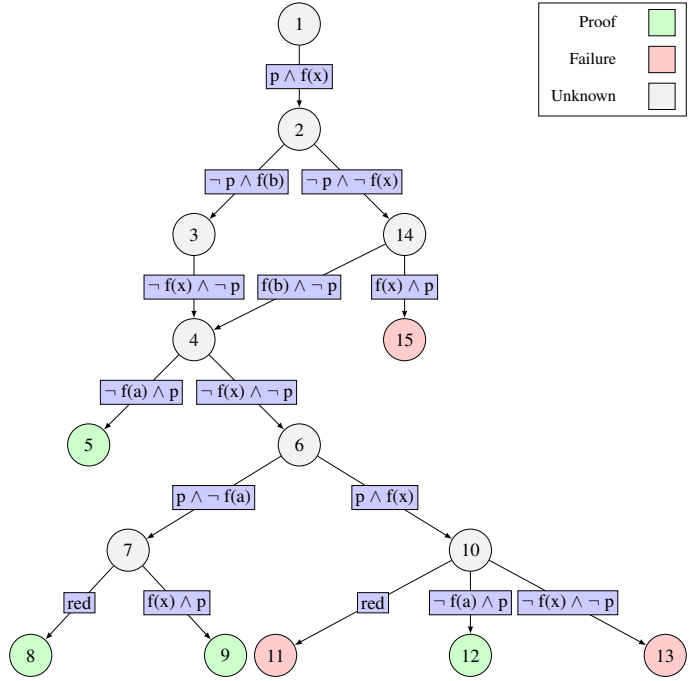
\begin{figure*}[ht]
  \centering
  \scalebox{0.7}{
  \begin{tikzpicture}[
    node/.style={circle, draw, fill=gray!10, minimum size=8mm, inner sep=0pt},
    edge label/.style={rectangle, draw, fill=blue!20, inner sep=2pt},
    level 1/.style={sibling distance=2.5cm},
    level 2/.style={sibling distance=4.0cm},
    level 3/.style={sibling distance=2.5cm},
    level 4/.style={sibling distance=4.0cm},
    level 5/.style={sibling distance=6.5cm},
    level 6/.style={sibling distance=3.7cm},
    level distance=2.0cm,
    edge from parent/.style={draw, -latex}
    ]

  \node[node] {1}
    child {node[node] {2}
      child {node[node] {3}
        child {node[node](4) {4}
          child {node[node, fill=green!20](5) {5}
          edge from parent
          node[edge label] {$\neg$ f(a) $\land$ p}}
          child {node[node](6) {6}
            child {node[node](7) {7}
              child {node[node, fill=green!20](8) {8}
              edge from parent
              node[edge label] {red}}
              child {node[node, fill=green!20](9) {9}
              edge from parent
              node[edge label] {f(x) $\land$ p}}
            edge from parent
            node[edge label] {p $\land$ $\neg$ f(a)}}
            child {node[node](10) {10}
              child {node[node, fill=red!20](11) {11}
              edge from parent
              node[edge label] {red}}
              child {node[node, fill=green!20](12) {12}
              edge from parent
              node[edge label] {$\neg$ f(a) $\land$ p}}
              child {node[node, fill=red!20](13) {13}
              edge from parent
              node[edge label] {$\neg$ f(x) $\land$ $\neg$ p}}
            edge from parent
            node[edge label] {p $\land$ f(x)}}
          edge from parent
          node[edge label] {$\neg$ f(x) $\land$ $\neg$ p}}
        edge from parent
        node[edge label] {$\neg$ f(x) $\land$ $\neg$ p}}
      edge from parent
      node[edge label] {$\neg$ p $\land$ f(b)}}
      child {node[node](14) {14}
        child {node[node, fill=red!20](15) {15}
        edge from parent
        node[edge label] {f(x) $\land$ p}}
      edge from parent
      node[edge label] {$\neg$ p $\land$ $\neg$ f(x)}}
    edge from parent
    node[edge label] {p $\land$ f(x)}};
    \draw[-latex] (14) -- node[edge label] {f(b) $\land$ $\neg$ p} (4);

    \begin{scope}[overlay]
      \node[rectangle, draw, fill=white, align=left, anchor=north east] at (current bounding box.north east) {
        \begin{tikzpicture}[
          legend node/.style={rectangle, draw, minimum size=5mm, inner sep=2pt, font=\small},
          legend text/.style={font=\small, align=left}
          ]

          \node[legend node, fill=green!20] (n1) at (1.0,0) {};
          \node[legend text] (t1) at (0,0) {Proof};
  
          \node[legend node, fill=red!20] (n2) at (1.0,-0.7) {};
          \node[legend text] (t2) at (0,-0.7) {Failure};
  
          \node[legend node, fill=gray!10] (n3) at (1.0,-1.4) {};
          \node[legend text] (t3) at (0,-1.4) {Unknown};
        \end{tikzpicture}
      };
    \end{scope}
  \end{tikzpicture}
  }
  \caption{Complete \lc search DAG of a simple theorem. Each node
    represents a \lc tableau, which is a directed tree whose nodes are
    labelled by first order literals. Tableau trees and further
    details are provided in Appendix~\ref{app:tableau}.}
  \label{fig:search_dag}
\end{figure*}

We propose to abstract away from the particularities of the proof
search and focus on the discovered, raw theorem proving data. We
disregard MCTS statistics and consider the explored tree as a set of
derivations, that can be assigned into the categories described in
Section~\ref{sec:pll_guidance}: 1) proofs, 2) failures and 3)
unknowns. We illustrate this in Figure~\ref{fig:search_dag}, showing
the complete search graph of a simple problem.

As a result, we obtain a \emph{partial} PLL dataset
$\dataset_{\textrm{MCTS}} =\big\{\big(\myinput^{(j)},
\mylabel^{(j)}\big)\big\}_{j=1}^{\nsamples}$ where \emph{partial}
refers to the fact that $\mylabel^{(j)}$ only identifies a handful of
the infinitely many possible derivations of $\myinput^{(j)}$.

In Section~\ref{sec:experiments} we compare different PLL methods for
training policy and evaluate them via the guided MCTS of \plc.

\section{PLL Methods for Policy Training}
\label{sec:pll_methods}

In this section, we give an overview of recent successful loss
fuctions for PLL, based on \citet{libra}. We make no attempt at giving
a complete coverage and we will not try to identify a single best
method for theorem proving. Instead, we intend to highlight the range
of approaches available for PLL based theorem proving. Each of these
methods assumes that we can compute the model predicted probabilities
of allowed labels. The optimisation objective is a function of these
probabilities.


\paragraph{Negative Log Likelihood loss (NLL)}
The $\nllloss$ is the standard example of an average-based loss that
appears frequently in the PLL literature, often under different
names. For example, it is called the \emph{maximum marginal likelihood
(MML)} loss in \citet{guu-etal} and the \emph{classifier consistent
(CC)} loss in \citet{provablyconsistentpll}.
$$
\Loss_{\nll}\left(\probs, \mylabel \right) 
= - \log \bigg( \sum_i \mylabelscalar_i \probsscalar_i \bigg)
= - \log(\pacc)
$$
where $\pacc$ is the predicted probability of all acceptable labels
$\sum_{i=1}^{\outdim} \probsscalar_i \mylabel_i$. The $\nllloss$ is
shown to have a strong bias in \citet{libra}, called the
\emph{winner-take-all} behaviour, often resulting in a degenerate
distribution where all the probability mass is placed on a single
allowed label with highest initial probability.
        
\paragraph{Uniform loss}
A very simple baseline is to compute the negative log likelihood of each allowed output and optimize their sum:
$$
\Loss_u(\probs, \mylabel)  = - \sum_i \mylabelscalar_i \log(\probsscalar_i)
$$
This is an average-based method which we refer to as
\emph{$\uloss$}. This is the loss used implicitly when different
proofs of the same problem are treated as independent samples and the
log probability of each is maximised. This loss has a single minimum,
when the prediction is uniform on the allowed outputs and zero
elsewhere.

\paragraph{$\beta$-Meritocratic loss}
\citet{guu-etal} proposes $\bmeritloss$, an identification-based
approach, which tries to model the likelihood that allowed label $\mylabelscalar_i$ is
the true label via weight $w(\beta)_i = \frac{\mylabelscalar_i
  \cdot \left( \probsscalar_i / \pacc \right)^\beta}{\sum_q
  \mylabelscalar_q \cdot \left( \probsscalar_q / \pacc
  \right)^\beta}$:
$$
\Loss_{\bmerit}(\probs, \mylabel) = - \sum_i w(\beta)_i \log(\probsscalar_i).
$$
The dependence of the $w(\beta)_i$ weights on the model output is
disregarded during optimisation, i.e., no gradients are propagated
through them. The $\beta$ parameter provides one possible smooth
interpolation between $\nllloss$ and $\uloss$: $\bmeritloss$ has the
same gradient as $\nllloss$ when $\beta=1$ and the gradient is
equivalent to that of $\uloss$ when $\beta=0$.  Based on a small grid
search, we use $\beta=0.5$ in our experiments and refer to this as
$\bmeritlossB$.

\paragraph{Libra-loss}
A recent loss function proposed in \citet{libra} is called the $\prploss$:
$$
\Loss_{\prplosssubscript}(\probs, \mylabel) = 
\underbrace{-\frac{1}{k} \sum_{i=1}^m \mylabelscalar_i \log(\probsscalar_i)}_{\textup{Allowed term}}
+ \underbrace{\log \bigg(1- \sum_{i=1}^m \mylabelscalar_i \probsscalar_i \bigg)}_{\textup{Disallowed term}}
$$ where $\numallowed = \sum_i \mylabelscalar_i$ is the number of
allowed outputs. The first term is the average of individual negative
log likelihood losses for each allowed output and only differs from
the $\uloss$ by a multiplicative factor of
$\frac{1}{\numallowed}$. The second term is the positive log
likelihood of selecting a disallowed label.  $\prploss$ was designed
to avoid the winner-take-all behaviour of $\nllloss$.  \citet{libra}
shows that this loss can be interpreted as a generalization of
$\nllloss$ extended with an input-dependent entropy regularization. In
each step this loss aims to increase the joint probability of
the allowed labels $\pacc$ while also preserving the probability
ratios among allowed labels.





\subsection{Comparing Loss Functions}
$\uloss$ represents one extreme because it aims towards uniform
distribution among alternatives, i.e., it forces each proof to have
the same predicted probability. $\nllloss$ represents the other
extreme: training dynamics tend to produce deterministic solutions,
i.e., the trained model tends to select a single proof and maximise
its probability. $\bmeritloss$ interpolates between the two in the
gradient space. $\prploss$ can also be seen as an interpolation
between the two extremes, but this is done differently for each input,
with the intention to avoid large changes to the probability ratio
among alternative proofs. See \citet{libra} for more analysis on the
loss functions.

\section{Experiments}
\label{sec:experiments}

The loss functions described in the previous section can be applied in
any offline learning setup when the policy model is updated based on a
large collection of proofs found. Here, we use the latest variant of
the \plc theorem prover to conduct experiments.

In order to ensure compability with previous work, we employ the same
graph neural network (GNN) model used in
\citet{prop_invariant_embedding} and \citet{plcop_gnn}, designed
specifically for logic. Since the previous codebase has become
obsolete\footnote{The old GNN code uses Tensorflow~1, which is not
supported any more by current hardware.}, we use our own
reimplementation of the GNN based learner in PyTorch.  In this paper
we do not address the question of best learning architecture, and
merely focus on how learning data is used.


\subsection{Datasets and Common Settings}
\label{subsec:datasets}
We experiment with ATP benchmarks that have been used
extensively in related systems.
The \emph{M2K} dataset \citep{m2k} is an extract of the larger Mizar40
dataset~\citep{mizar40} and consists of 2003 problems that come from
related Mizar articles. The bushy variant from the \emph{MPTP2078}
benchmark~\citep{abs-1108-3446} contains an article-based
selection of Mizar problems.
Finally we make use of the
\emph{RA-2} dataset first used in \citet{flop}, which consists of 1000
problems in the theory of Robinson Arithmetic. Each theorem is an
equation of the form $T = N$, where $0 \leq N$, and $T$ is a random
expression with $3$ operators and operands $N_i$ such that $0 \leq N_i
< 10.$ (E.g.: $((3\!+\!4)\!\cdot\!2)\!+\!6\!=\!20$.)

The previous version of \plc, referred to as \textbf{pGNN}, trains the
model to imitate nodewise MCTS statistics. As a sanity check, we
reimplement the associated loss function and use it as our baseline,
referring to it as \textbf{BS}. Note, however, that differences remain
in the grouping of training samples. pGNN mixes nodes across problems
and draws a random batch of nodes for each gradient update. In BS, we
focus on loss extracted from one single and entire MCTS tree. This
gives some rigidity to the new system when learning to imitate MCTS
statistics, but it is more suitable for the other loss functions that
we examine. Furthermore, to speed up experimentation, we make several
simplifications. First, we reduce the number of message passing layers
in the model from 10 to 5. Second, we reduce the number of training
epochs from 100 to 10. Finally, while pGNN uses separate models for
training the value and the policy, we use a single model with separate
heads. Any other hyperparameters are the same as described in
\citet{plcop_gnn}. In particular, we do not alter the value learning
mechanism and only focus on the policy.

Table~\ref{tab:m2k_previous} compares BS with pGNN on the M2K
dataset. BS surpasses pGNN in the first 4-5 iterations, but it
saturates afterwards, likely due to the smaller model size, shorter
training and the more rigid batching mechanism. 

\begin{table}[t]
  \caption{Number of problems solved on the M2k dataset by pGNN and BS.}
  \label{tab:m2k_previous}
  \centering
  \addtolength{\tabcolsep}{-0.3em}
  \begin{tabular}{ l | l l l l l l l}
    Iteration & 0 & 1 & 2 & 4 & 6 & 8 & 10 \\
    \toprule
    pGNN & 548 & 727  & 854  & 1057 & \textbf{1132} & \textbf{1184} & \textbf{1228} \\
    BS       & 548 & \textbf{934}  & \textbf{1037} & \textbf{1061} & 1071 & 1059 & 1116 \\
    \bottomrule
  \end{tabular}
  \addtolength{\tabcolsep}{0.3em}
\end{table}

We highlight the fact that our approach yields expensive
experiments. While previous works use single MCTS nodes as training
samples, for us an entire MCTS search tree constitutes a single
sample. Consequently, a single update might cost as much as thousands
of updates in other systems, depending on the size of the search
tree. For illustration, a single experiment with 5 iterations of
learning on the MPTP2078 dataset takes around 8 days with an NVIDIA
GeForce RTX 2080 GPU for training and with 40 cores for parallel MCTS.
This can be mitigated by extensive engineering,
however, here we only resort to
demonstrating the potential of taking a global view on the entire
search tree. As a result, we analyse the models' sensitivity to
initial configurations only on some shorter experiments.

\subsection{Experiment 1: Training on all Proofs}

\begin{figure*}[htb]
  \begin{center}
    \includegraphics[width=0.48\linewidth]{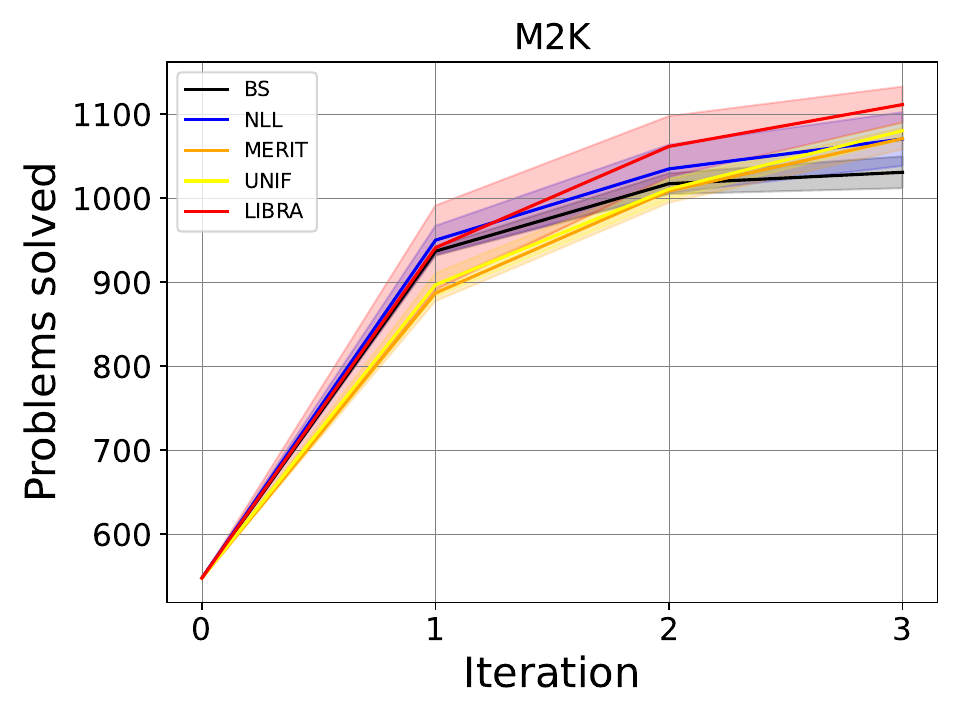}
    \includegraphics[width=0.48\linewidth]{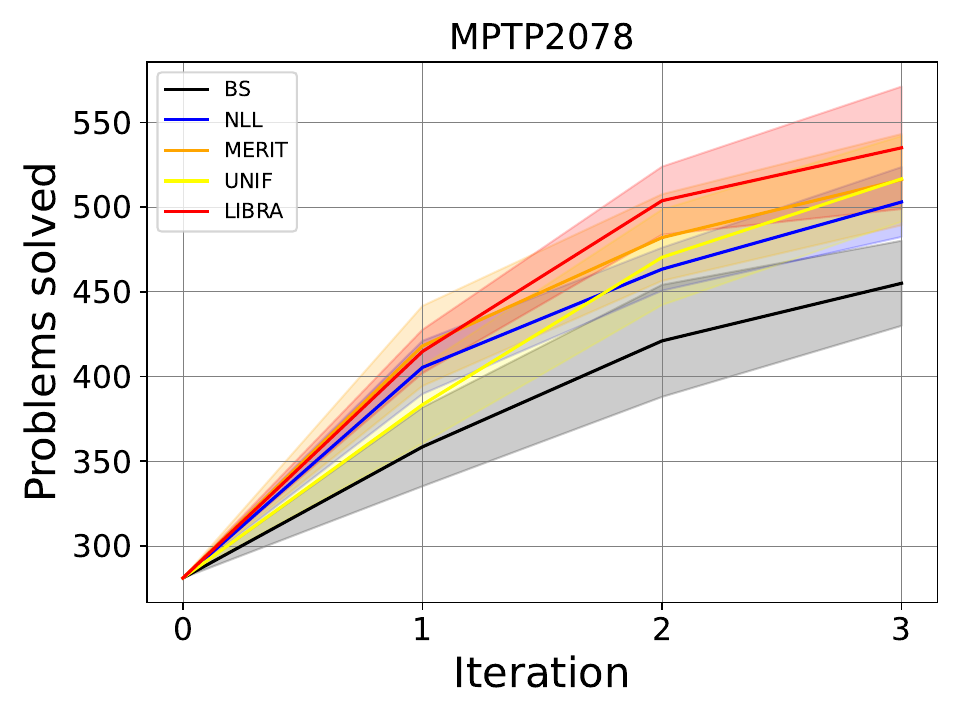}
    \caption{Number of problems solved on the M2k and MPTP dataset by guided MCTS. The curves show averages based on 5 trials, Shaded areas indicate standard variance.}
    \label{fig:variance}
  \end{center}
\end{figure*}

We contrast BS with methods from PLL that exploit all available proofs
during training.  We perform 3 iterations of guided MCTS on the M2K
and MPTP2078 datasets using BS, $\nllloss$ (NLL), $\bmeritlossB$
(MERIT), $\uloss$ (UNIF) and $\prploss$
(LIBRA). Figure~\ref{fig:variance} shows the average number of problems
solved, along with standard deviation, based on 5 trials.

The figure shows a large variance and it is not easy to identify
a clear ordering among the different methods. However, LIBRA tends to
perform best and BS tends to perform worse. Overall, there is a
significant benefit of using one of the PLL methods instead of BS.
We take the best performing models and train them for two additional iterations. The results are provided in Table~\ref{tab:trainmore}.
We find an improvement of 14-28\% over the baseline BS. The
best PLL variants seem to be $\prploss$ and $\bmeritlossB$, both of
which were designed to maintain a balance among alternative proofs.

\begin{table*}[t]
  \caption{Number of problems solved on the M2K and MPTP2078 datasets over 5
    iterations of guided MCTS.}
  \label{tab:trainmore}
  \centering
  \begin{tabular}{ l | r r r r r r || r r r r r r }
    Iteration & 0 & 1 & 2 & 3 & 4 & 5 & 0 & 1 & 2 & 3 & 4 & 5 \\
    \toprule
    Dataset & \multicolumn{6}{c||}{M2K} & \multicolumn{6}{c}{MPTP2078} \\
    \midrule
    BS        & 548 & 934 & 1037 & 1007 & 1061 & 1059 & 280 & 374 & 448 & 454 & 454 & 507 \\
    \midrule
    NLL       & 548 & 941 & 1039 & 1085 & 1105 & 1136 & 280 & 402 & 468 & 510 & 558 & 566 \\
    MERIT     & 548 & 998 & 1074 & 1129 & 1153 & 1183 & 280 & \textbf{449} & \textbf{509} & 554 & \textbf{621} & \textbf{647} \\
    UNIF      & 548 & 917 & 1017 & 1077 & 1135 & 1170 & 280 & 399 & 499 & 486 & 554 & 513 \\
    LIBRA     & 548 & \textbf{1006} & \textbf{1116} & \textbf{1138} & \textbf{1186} & \textbf{1205} & 280 & 408 & 486 & \textbf{556} & 591 & 620 \\
    \bottomrule
  \end{tabular}
\end{table*}

\subsection{Experiment 2: Understanding the Effect of Alternative Proofs}

The RA-2 dataset is rather peculiar in that MCTS barely ever finds
alternatives. After the unguided iteration, 80 problems are
solved out of 1000 and only a single problem has two proofs and no
problem has more than 2 proofs. In this setup, it is not clear if the
PLL variants make any sense, since there are no alternatives to
balance among. Nevertheless, Table~\ref{tab:arithmetic_pll} shows that
three of our four PLL methods show very strong performance, surpassing
the baseline by $19-24\%$. Surprisingly, $\nllloss$ exhibits very weak learning.

\begin{table}[htb]
  \caption{Number of problems solved on the RA-2 dataset, as well as
    the average number of proofs of solved problems (in parenthesis) for various
    iterations. The base model solves 80 problems.}
  \label{tab:arithmetic_pll}
  \centering

   \begin{tabular}{ l | r r r r r r }
    Iteration & 1 & 2 & 4 & 6 & 8 & 10 \\
    \toprule
    BS        & 78 (1.01) & 76 (2.1) & 150 (1.86) & 314 (1.84) & 460 (1.78) & 408 (1.82) \\
    NLL       & 91 (1.05) & 100 (1.02) & 131 (1.03) & 124 (1.05) & 128 (1.05) & 278 (1.05) \\
    MERIT     & 107 (1.01) & 117 (1.02) & 136 (1.02) & 337 (1.04) & 478 (1.05) & \textbf{505} (1.05) \\
    UNIF      & 106 (1.02) & 136 (1.14) & \textbf{440} (1.16) & \textbf{470} (1.16) & \textbf{487} (1.18) & 485 (1.18) \\
    LIBRA     & \textbf{114} (1.03) & \textbf{171} (1.11) & 241 (1.09) & 423 (1.08) & 483 (1.09) & 500 (1.08) \\
    \bottomrule
  \end{tabular}

\end{table}

Trying to understand the performance differences, we also show the
average number of proofs per solved problem. We see that BS generates
a greater number of proofs, which is expected as it is trained to
maintain the initial exploration/exploitation balance of MCTS. By
focusing more on increasing the probability of proofs found, the PLL
methods end up being more targeted, rarely finding more than one
proof. This works well for $\prploss$, $\bmeritlossB$ and $\uloss$,
while it harms $\nllloss$. We hypothesize that even if the
alternatives are few, it makes a big difference how probability is
distributed among them. $\nllloss$ is known to be weak at controlling
this balance and can get stuck at unwanted local minima, while the
others can better adapt to training signal coming from other problems \citep{libra}.

MCTS has a hyperpameter $cp$, which is responsible for the
exploration/exploitation tradeoff: the higher its value, the more
effort MCTS will make to search for alternative proofs. In
Appendix~\ref{app:alternatives} we describe an experiment in which we
vary the number of proofs available for training. In that experiment,
we find that the number of proofs does not provide sufficient
information to compare PLL methods.

\subsection{Experiment 3: Training on a Single Proof}
One alternative of learning to imitate MCTS statistics is to pick a
single proof found and train the model to maximise its probability,
via minimising its negative log likelihood. Natural candidatates for
the single proof are the shortest, the longest or a uniform
sample. Alternatively, one could also consider training on failed
derivations as well: in which case the model is trained to avoid such
derivations. In Appendix~\ref{app:singleproof} we describe
experimental results that suggest that the shortest proof tends to
work best, however, it performs weaker than the PLL methods.

\subsection{Experiment 4: Comparison with Previous Works}

M2K is the dataset most extensively used in previous experiments with
guided MCTS. For this reason we compare BS and $\prploss$ with
published results. To make the comparison fairer, we increase
the training time from 10 to 100 epochs and train separate models for
policy and value, as done in \citet{plcop_gnn}.
The results are
summarised in Table~\ref{tab:m2k_previous2}.

\begin{table}[t]
  \caption{Number of problems solved on the M2k dataset by 
    baseline models. \rlc and \plc XGB use 10 times as much inference
    budget than later variants. {\bf Incr} shows the performance increase in percentages from
    the unguided model to the best result.}
  \label{tab:m2k_previous2}
  \centering
  \addtolength{\tabcolsep}{-0.4em}
  \begin{tabular}{ l | l l l l l l l | l}
    Iteration & 0 & 1 & 2 & 4 & 6 & 8 & 10 & Incr \\
    \toprule
    \rlc  (10)        & 770 & 1037 & 1110 & 1179 & 1198 & 1193 & 1210
    & $57\%$ \\
    pXGB (10)      & 695 & 1070 & 1209 & 1295 & 1322 & 1339 &
    1359 & $96\%$ \\
    pXGB      & 548 & 790  & 956  & 1061 & 1119 & 1147 & 1171 & $114\%$ \\
    pGNN      & 548 & 727  & 854  & 1057 & 1132 & 1184 & 1228 & $124\%$\\
    BS        & 548 & 826  & 981 & 1080 & 1079 & 1102 & 1119 & $104\%$\\
    LIBRA & 548 & 981 & 1133 & 1205 & 1273 & 1294 & 1313 & $140\%$\\
    \bottomrule
  \end{tabular}
  \addtolength{\tabcolsep}{0.4em}
\end{table}

The rows in the table follow the order of publication. \rlc is the
first system and relies on fast, simple learners
\citep[XGBoost][]{xgboost} and MCTS is given a larger inference budget
of 200000 steps. Later rows correspond to developing variants of
\plc. \textbf{pXGB} keeps the models while introducing enhancements to
the base calculus and the data extraction. Later systems use GNN based
guidance, which is more powerful, but slower. Consequently, these
systems use a reduced inference budget of only 20000 steps to avoid
being overly slowed down, while still being capable of strong
learning. For pXGB, we show its performance both with the small and
the large step limit.


After 10 iterations of training, $\prploss$ outperforms BS by $17\%$
and pGNN by $7\%$. $\prploss$ is minimally weaker than pXGB (by
$2\%$) which uses 10 times as much inference and if we consider the
improvement ratio with respect to the unguided baseline, $\prploss$
turns out to be the best.

\section{Discussion and Conclusion}
\label{sec:conclusion}

Our paper provides a systematic study of learning from alternative
proofs to guide an ATP system and shows that PLL provides a suitable
theoretical framework for tackling this problem.


We find that a consistent improvement can be achieved in the \plc
system via PLL methods that learn jointly from all proofs, surpassing
both the baseline which learns to imitate statistics of the Monte
Carlo Search Tree and single proof strategies.
We highlight that
on the M2K dataset $\prploss$ yields $7\%$ improvement over the
previous best result with guided MCTS on the \lc prover.

One of the major questions that remain to be answered is how to choose
among different PLL methods for different datasets. We have made some
probing experiments to relate the number of alternative proofs with
the prover's performance, but we found no satisfying conclusion.

Our work can be extended in numerous directions to better understand
the scope of improvement. Other datasets, other proving calculi, other
PLL methods can be tried and this requires no changes to the prover or
the learner architecture: only the data collection and the loss
function need to be updated.






\begin{ack}
This work has been supported by the the European Union project RRF-2.3.1-21-2022-00004
within the framework of the Artificial Intelligence National Laboratory Grant
no RRF-2.3.1-21-2022-00004, as well as the ELTE TKP 2021-NKTA-62 funding scheme.
\end{ack}

\bibliography{manyproofs}
\bibliographystyle{plainnat}

\newpage
\appendix

\section{Training on a Single Proof}
\label{app:singleproof}

One alternative of learning to imitate MCTS statistics is to pick a
single proof found and train the model to maximise its probability,
via minimising its negative log likelihood. Natural candidatates for
the single proof are the shortest (SHORT), the longest (LONG) or a
uniform sample (RAND). The first section of Table~\ref{tab:bigcompare}
shows performance for these choices on the M2K and MPTP2078 datasets.

\begin{table*}[htb]
  \caption{Number of problems solved on the M2K and MPTP2078 datasets over 5
    iterations of guided MCTS.
  }
  \label{tab:bigcompare}
  \centering
  \begin{tabular}{ l | r r r r r r || r r r r r r }
    Iteration & 0 & 1 & 2 & 3 & 4 & 5 & 0 & 1 & 2 & 3 & 4 & 5 \\
    \toprule
    Dataset & \multicolumn{6}{c||}{M2K} & \multicolumn{6}{c}{MPTP2078} \\
    \midrule
    BS        & 548 &  934 & 1037 & 1007 & 1061 & 1059 & 280 & 374 & 448 & 454 & 454 & 507 \\
    \midrule
    SHORT     & 548 &  935 & 1072 & 1114 & 1153 & 1175 & 280 & 404 & 478 & 473 & 509 & 547 \\
    LONG      & 548 &  875 &  997 & 1055 & 1090 & 1128 & 280 & 388 & 442 & 529 & 545 & 541 \\
    RAND      & 548 &  866 &  991 & 1067 & 1103 & 1141 & 280 & 402 & 429 & 495 & 524 & 550 \\
    \midrule
    SHORT $\pm$ & 548 &  892 & 1036 & 1101 & 1144 & 1168 & 280 & 406 & 449 & 541 & 583 & 591 \\
    LONG $\pm$  & 548 &  828 & 968  & 1024 & 1066 & 1109 & 280 & 399 & 460 & 468 & 505 & 545 \\
    RAND $\pm$  & 548 &  862 & 1010 & 1064 & 1110 & 1127 & 280 & 435 & 494 & 477 & 540 & 574 \\
    \midrule
    NLL       & 548 & 941 & 1039 & 1085 & 1105 & 1136 & 280 & 402 & 468 & 510 & 558 & 566 \\
    MERIT     & 548 & 998 & 1074 & 1129 & 1153 & 1183 & 280 & \textbf{449} & \textbf{509} & 554 & \textbf{621} & \textbf{647} \\
    UNIF      & 548 & 917 & 1017 & 1077 & 1135 & 1170 & 280 & 399 & 499 & 486 & 554 & 513 \\
    LIBRA     & 548 & \textbf{1006} & \textbf{1116} & \textbf{1138} & \textbf{1186} & \textbf{1205} & 280 & 408 & 486 & \textbf{556} & 591 & 620 \\
    \bottomrule
  \end{tabular}
\end{table*}

We find that training on a single proof brings a consistent 5-10\%
improvement over BS, while it performs weaker than the PLL
variants. The shortest proof tends to work best and it outperforms BS
in all experiments, confirming previous results in other proving
systems. Note, futhermore, that random choice is also very competitive.

MCTS does not only provide the learner with proofs, but also with failed
derivations. We consider training on pairs of derivations: a proof
to imitate and a failed derivation to avoid. For proofs we maximise
the log probability of the derivation, as in the previous
experiment. For failures, we simply minimise their log
probability. Again, we can consider shortest derivations (SHORT
$\pm$), longest derivations (LONG $\pm$) and random choice (RAND
$\pm$). The training results can be seen in the second section of
Table~\ref{tab:bigcompare}.  Overall, we see that the presence of
failed derivations slightly harms performance. The only exception is
MPTP2078 where adding the shortest failed derivation to the shortest
proof brings 8\% improvement (591 vs 547).

The negative effect of failed derivations could potentially be due to
the fact that there are so many possible failures and learning from a
single one might not provide useful signal. Another possible source of
the problem could be the lack of proper tuning a hyperparameter that
strikes the right balance among the positive and negative loss
terms. We leave this question for future work and only conclude that
incorporating failed derivations does not seem to lead to easy
improvement.

\section{Changing the Number of Alternative Proofs to Train from}
\label{app:alternatives}

We aim to get a better understanding of how the number of alternative
proofs influences different training methods. We vary the $cp$
hyperparameter of MCTS, which is responsible for the
exploration/exploitation tradeoff: the higher its value, the more
effort MCTS will make to search for alternative
proofs. Table~\ref{tab:cp} shows for the M2K and
MPTP2078 datasets how this influences the number of proofs per
problem used for training, as well as the number of problems solved in
the first guided iteration.

We see that changing $cp$ indeed greatly changes the number of proofs
per problem. However, we also see that having much more alternatives
does not help the trainer and performance usually peaks at around $cp
\in \{1,2\}$, which coincides with the highest number of solved
problems. Another interesting observation is that BS is much more
sensitive to changing the number of proofs, as shown by the standard
deviation across different $cp$ values. However, what we do not see in
this table is any explanation about why one method might perform
better than another. Overall, we conclude that the number of proofs
does not provide sufficient information to compare PLL methods.

\begin{table}[htb]
  \caption{Number of problems solved on the M2K and MPTP2078 datasets
    in the first guided iteration for different $cp_0$ values of the
    unguided iteration. \emph{solved$_0$} and \emph{proofs$_0$} give
    the number of problems solved and the average number of proofs per
    solved problem in the unguided iteration. \emph{Std} refers to the
    standard deviation across different $cp_0$ values. \textbf{Bold}
    text indicates which $cp_0$ value is best for the given method.}
  \label{tab:cp}
  \centering
  \addtolength{\tabcolsep}{-0.2em}
  \begin{tabular}{ l | r r r r r r | r || r r r r r r | r}
    $cp_0$ & 0.5 & 1 & 2 & 5 & 10 & 50 & Std & 0.5 & 1 & 2 & 5 & 10 & 50 & Std \\
    \toprule
    Dataset & \multicolumn{7}{c||}{M2K} & \multicolumn{7}{c}{MPTP2078} \\
    \midrule
    solved$_0$ & 469 & 558 & 548 & 516 & 496 & 482 & 33 & 230 & 263 & 281 & 270 & 251 & 250 & 16 \\
    proofs$_0$ & 1.0 & 1.2 & 2.6 & 7.9 & 13.8 & 29.4 & & 1.0 & 1.1 & 1.8 & 8.1 & 16.4 & 45.0 &  \\
    \toprule
    BS        & 819 & 889 & \textbf{934} & 816 & 656 & 597 & 120 & 302 & 366 & \textbf{374} & 333 & 293 & 275 & 37 \\
    \midrule
    SHORT   & 909 & \textbf{977} & 935 & 906 & 876 & 784 & 60 & 357 & 396 & \textbf{404} & 375 & 385 & 384 & 15 \\
    LONG    & 888 & \textbf{947} & 875 & 876 & 837 & 768 & 54 & 361 & 292 & 388 & \textbf{404} & 362 & 377 & 35 \\
    RAND    & 879 & 886 & 866 & 874 & \textbf{903} & 822 & 25 & 346 & 381 & \textbf{402} & 366 & 365 & 350 & 19 \\
    \midrule
    SHORT $\pm$ & 748 & 842 & 892 & \textbf{939} & 798 & 829 & 62 & 279 & 394 & \textbf{406} & 318 & 336 & 333 & 44 \\
    LONG $\pm$  & 909 & \textbf{967} & 828 & 850 & 891 & 853 & 46 & 403 & \textbf{410} & 399 & 348 & 369 & 401 & 22 \\
    RAND $\pm$  & 841 & 904 & 862 & \textbf{929} & 883 & 805 & 41 & 396 & 416 & \textbf{435} & 393 & 371 & 398 & 20 \\
    \midrule
    NLL       & 914 & 938 & \textbf{941} & 927 & 897 & 894 & 18 & 382 & \textbf{425} & 402 & 330 & 392 & 352 & 31 \\
    MERIT     & 924 & 956 & 872 & \textbf{960} & 884 & 892 & 34 & 418 & 406 & \textbf{449} & 375 & 411 & 428 & 22 \\
    UNIF      & 911 & \textbf{953} & 917 & 825 & 813 & 742 & 73 & 354 & \textbf{413} & 399 & 374 & 364 & 343 & 25 \\
    LIBRA     & 971 & 954 & \textbf{1006} & 867 & 896 & 836 & 60 & 408 & 417 & 414 & 363 & \textbf{426} & 401 & 20 \\
    \bottomrule
  \end{tabular}
  \addtolength{\tabcolsep}{0.2em}
\end{table}

\section{Tuning the Number of Message Passing Layers in the GNN}
\label{sec:traintimeandloss}

\begin{figure}[ht]
  \begin{center}
\begin{tikzpicture}
  \begin{axis}[
      xlabel={Epoch},
      ylabel={Proof Probability},
      xmin=0, xmax=42,
      ymin=0, ymax=0.65,
      width=0.7\textwidth,
      height=0.3\textheight,
      legend style={at={(1.05,1)}, anchor=north west, font=\small},
      ymajorgrids=true,
      grid style=dashed,
  ]

  \addplot[
      color=blue,
      solid
      ]
      coordinates {
          (0,0.089) (1,0.186) (2,0.215) (3,0.225) (4,0.234) (5,0.240)
          (6,0.245) (7,0.244) (8,0.249) (9,0.255) (10,0.258) (11,0.263)
          (12,0.272) (13,0.271) (14,0.279) (15,0.287) (16,0.286) (17,0.300)
      };
  \addlegendentry{Layer 3 - Train}

  \addplot[
      color=blue,
      dashed
      ]
      coordinates {
          (0,0.089) (1,0.174) (2,0.216) (3,0.239) (4,0.282) (5,0.282)
          (6,0.271) (7,0.272) (8,0.272) (9,0.247) (10,0.276) (11,0.253)
          (12,0.241) (13,0.257) (14,0.298) (15,0.216) (16,0.254) (17,0.285)
      };
  \addlegendentry{Layer 3 - Validation}

  \addplot[
      color=red,
      solid
      ]
      coordinates {
          (0,0.089) (1,0.185) (2,0.222) (3,0.239) (4,0.251) (5,0.257)
          (6,0.263) (7,0.258) (8,0.264) (9,0.283) (10,0.285) (11,0.287)
          (12,0.296) (13,0.303) (14,0.304) (15,0.322) (16,0.332) (17,0.350)
          (18,0.349) (19,0.365) (20,0.379) (21,0.396) (22,0.421) (23,0.435)
          (24,0.451) (25,0.464)
      };
  \addlegendentry{Layer 4 - Train}

  \addplot[
      color=red,
      dashed
      ]
      coordinates {
          (0,0.089) (1,0.216) (2,0.186) (3,0.264) (4,0.289) (5,0.308)
          (6,0.244) (7,0.287) (8,0.282) (9,0.288) (10,0.313) (11,0.279)
          (12,0.294) (13,0.271) (14,0.326) (15,0.314) (16,0.262) (17,0.286)
          (18,0.300) (19,0.329) (20,0.332) (21,0.321) (22,0.326) (23,0.308)
          (24,0.324) (25,0.344)
      };
  \addlegendentry{Layer 4 - Validation}

\addplot[
  color=green,
  solid
  ]
  coordinates {
      (0,0.089) (1,0.184) (2,0.216) (3,0.228) (4,0.235)
      (5,0.248) (6,0.257) (7,0.258) (8,0.267) (9,0.283)
      (10,0.288) (11,0.298) (12,0.300) (13,0.323) (14,0.328)
      (15,0.341) (16,0.358) (17,0.370) (18,0.383) (19,0.395)
      (20,0.421) (21,0.447) (22,0.457) (23,0.473) (24,0.487)
      (25,0.530) (26,0.543)
  };
\addlegendentry{Layer 5 - Train}

\addplot[
  color=green,
  dashed
  ]
  coordinates {
      (0,0.089) (1,0.205) (2,0.179) (3,0.231) (4,0.298)
      (5,0.298) (6,0.252) (7,0.279) (8,0.286) (9,0.269)
      (10,0.324) (11,0.284) (12,0.265) (13,0.279) (14,0.337)
      (15,0.278) (16,0.313) (17,0.302) (18,0.334) (19,0.333)
      (20,0.353) (21,0.352) (22,0.337) (23,0.328) (24,0.353)
      (25,0.360) (26,0.347)
  };
\addlegendentry{Layer 5 - Validation}

\addplot[
    color=purple,
    solid
    ]
    coordinates {
        (0,0.088) (1,0.178) (2,0.216) (3,0.230) (4,0.241)
        (5,0.260) (6,0.255) (7,0.254) (8,0.264) (9,0.281)
        (10,0.278) (11,0.282) (12,0.286) (13,0.296) (14,0.307)
        (15,0.315) (16,0.321) (17,0.330) (18,0.349) (19,0.353)
        (20,0.373) (21,0.379) (22,0.415) (23,0.435) (24,0.440)
        (25,0.467) (26,0.492) (27,0.488)
    };
\addlegendentry{Layer 6 - Train}

\addplot[
    color=purple,
    dashed
    ]
    coordinates {
        (0,0.088) (1,0.159) (2,0.175) (3,0.270) (4,0.301)
        (5,0.295) (6,0.258) (7,0.266) (8,0.279) (9,0.262)
        (10,0.303) (11,0.285) (12,0.260) (13,0.305) (14,0.305)
        (15,0.267) (16,0.287) (17,0.291) (18,0.320) (19,0.326)
        (20,0.351) (21,0.318) (22,0.325) (23,0.304) (24,0.329)
        (25,0.332) (26,0.332) (27,0.333)
    };
\addlegendentry{Layer 6 - Validation}

\addplot[
    color=brown,
    solid
    ]
    coordinates {
        (0,0.088) (1,0.162) (2,0.201) (3,0.205) (4,0.225)
        (5,0.231) (6,0.233) (7,0.238) (8,0.246) (9,0.259)
        (10,0.258) (11,0.270) (12,0.272) (13,0.266) (14,0.282)
        (15,0.289) (16,0.288) (17,0.296) (18,0.307) (19,0.310)
        (20,0.328) (21,0.335) (22,0.355) (23,0.361) (24,0.377)
        (25,0.410) (26,0.424)
    };
\addlegendentry{Layer 7 - Train}

\addplot[
    color=brown,
    dashed
    ]
    coordinates {
        (0,0.088) (1,0.125) (2,0.183) (3,0.246) (4,0.287)
        (5,0.240) (6,0.246) (7,0.269) (8,0.270) (9,0.274)
        (10,0.297) (11,0.264) (12,0.241) (13,0.266) (14,0.298)
        (15,0.269) (16,0.259) (17,0.278) (18,0.313) (19,0.310)
        (20,0.327) (21,0.320) (22,0.331) (23,0.315) (24,0.314)
        (25,0.330) (26,0.314)
    };
\addlegendentry{Layer 7 - Validation}

\addplot[
    color=orange,
    solid
    ]
    coordinates {
        (0,0.089) (1,0.164) (2,0.201) (3,0.199) (4,0.205)
        (5,0.219) (6,0.216) (7,0.226) (8,0.235) (9,0.247)
        (10,0.237) (11,0.243) (12,0.250) (13,0.245) (14,0.252)
        (15,0.259) (16,0.260) (17,0.273) (18,0.270) (19,0.278)
        (20,0.291) (21,0.295) (22,0.295) (23,0.307) (24,0.306)
        (25,0.323) (26,0.329) (27,0.341) (28,0.344) (29,0.364)
        (30,0.374) (31,0.378) (32,0.401) (33,0.414) (34,0.430)
        (35,0.453) (36,0.464) (37,0.476)
    };
\addlegendentry{Layer 8 - Train}

\addplot[
    color=orange,
    dashed
    ]
    coordinates {
        (0,0.089) (1,0.258) (2,0.164) (3,0.261) (4,0.235)
        (5,0.242) (6,0.221) (7,0.256) (8,0.274) (9,0.272)
        (10,0.261) (11,0.231) (12,0.257) (13,0.260) (14,0.262)
        (15,0.250) (16,0.268) (17,0.284) (18,0.284) (19,0.272)
        (20,0.294) (21,0.317) (22,0.332) (23,0.277) (24,0.335)
        (25,0.327) (26,0.296) (27,0.306) (28,0.319) (29,0.345)
        (30,0.292) (31,0.310) (32,0.298) (33,0.316) (34,0.327)
        (35,0.307) (36,0.321) (37,0.332)
    };
\addlegendentry{Layer 8 - Validation}

\addplot[
    color=cyan,
    solid
    ]
    coordinates {
        (0,0.089) (1,0.149) (2,0.177) (3,0.202) (4,0.200)
        (5,0.199) (6,0.209) (7,0.213) (8,0.218) (9,0.229)
        (10,0.229) (11,0.230) (12,0.238) (13,0.241) (14,0.245)
        (15,0.259) (16,0.255) (17,0.267) (18,0.264) (19,0.268)
        (20,0.278) (21,0.281) (22,0.287) (23,0.288) (24,0.295)
        (25,0.304) (26,0.315) (27,0.311) (28,0.331) (29,0.340)
        (30,0.352) (31,0.359) (32,0.379) (33,0.374) (34,0.407)
        (35,0.428) (36,0.428) (37,0.463) (38,0.469)
    };
\addlegendentry{Layer 9 - Train}

\addplot[
    color=cyan,
    dashed
    ]
    coordinates {
        (0,0.089) (1,0.189) (2,0.183) (3,0.260) (4,0.176)
        (5,0.262) (6,0.202) (7,0.228) (8,0.269) (9,0.264)
        (10,0.272) (11,0.247) (12,0.232) (13,0.261) (14,0.288)
        (15,0.242) (16,0.264) (17,0.272) (18,0.298) (19,0.290)
        (20,0.295) (21,0.295) (22,0.304) (23,0.302) (24,0.296)
        (25,0.306) (26,0.285) (27,0.319) (28,0.318) (29,0.308)
        (30,0.312) (31,0.315) (32,0.306) (33,0.315) (34,0.330)
        (35,0.306) (36,0.325) (37,0.338) (38,0.360)
    };
\addlegendentry{Layer 9 - Validation}

\addplot[
    color=magenta,
    solid
    ]
    coordinates {
        (0,0.088) (1,0.162) (2,0.185) (3,0.197) (4,0.201)
        (5,0.207) (6,0.216) (7,0.189) (8,0.206) (9,0.222)
        (10,0.217) (11,0.232) (12,0.242) (13,0.239) (14,0.241)
        (15,0.240) (16,0.234) (17,0.246) (18,0.252) (19,0.260)
        (20,0.268) (21,0.264) (22,0.278) (23,0.280) (24,0.285)
        (25,0.299) (26,0.300) (27,0.300) (28,0.309) (29,0.321)
        (30,0.332) (31,0.339) (32,0.345) (33,0.348) (34,0.361)
        (35,0.386) (36,0.385) (37,0.402) (38,0.417) (39,0.439)
        (40,0.449) (41,0.473)
    };
\addlegendentry{Layer 10 - Train}

\addplot[
    color=magenta,
    dashed
    ]
    coordinates {
        (0,0.088) (1,0.156) (2,0.208) (3,0.252) (4,0.216)
        (5,0.257) (6,0.241) (7,0.245) (8,0.252) (9,0.269)
        (10,0.259) (11,0.262) (12,0.268) (13,0.263) (14,0.277)
        (15,0.246) (16,0.256) (17,0.256) (18,0.292) (19,0.264)
        (20,0.253) (21,0.293) (22,0.296) (23,0.258) (24,0.279)
        (25,0.305) (26,0.295) (27,0.306) (28,0.304) (29,0.325)
        (30,0.310) (31,0.288) (32,0.282) (33,0.302) (34,0.335)
        (35,0.293) (36,0.300) (37,0.332) (38,0.347) (39,0.342)
        (40,0.339) (41,0.323)
    };
\addlegendentry{Layer 10 - Validation}
  
  \end{axis}
\end{tikzpicture}
\caption{Aggregate model predicted probability of the proof derivations in the training data, for different number of message passing layers}
\label{fig:gnn_layers}
  \end{center}
\end{figure}
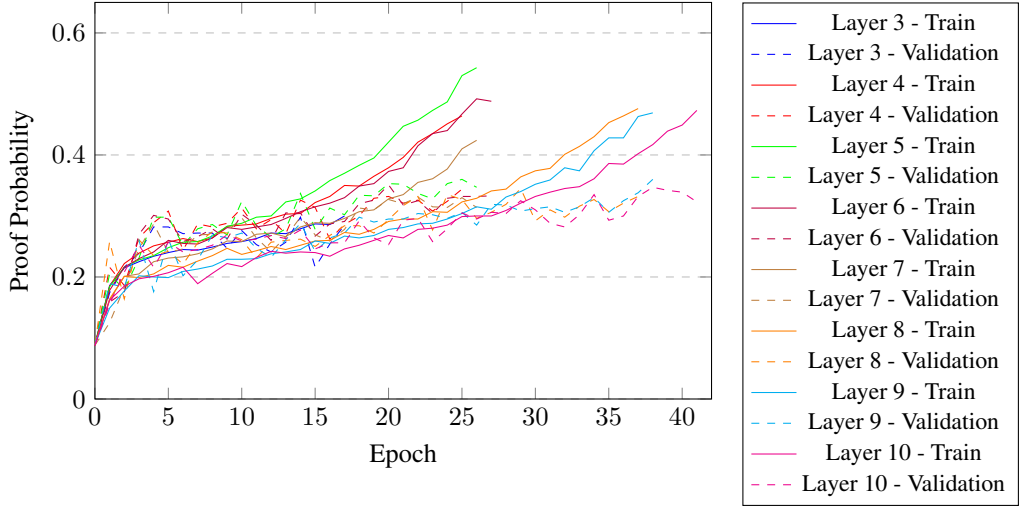

An important hyperparameter of our GNN model is the number of message
passing layers, which we tune based on the data extracted from the M2K
dataset by unguided MCTS. During training, our primary evaluation
metric is the sum of probabilities of proofs found for a given
problem. Figure~\ref{fig:gnn_layers} shows that increasing the number
of layers influences performance on the train set, but has little
effect on the validation set. For this reason, we run most experiments
with only 5 layers to speed up computation. Note however, that for
larger datasets and in later iterations, when there is more training
data available, extra layers could still prove beneficial.

\pagebreak
\section{Proof Search DAG and Tableau Trees}
\label{app:tableau}

For illustration, we provide the complete \lc search DAG of the proof of the following
simple statement:

\begin{equation}
  \exists x ((p \land \lnot f(a)) \lor (f(b) \land \lnot p) \lor (p
  \land f(x)) \lor (\lnot p \land \lnot f(x)))  \nonumber
\end{equation}

The above formula is in disjunctive normal form, i.e., it is a
disjunction of conjunctions of literals. If the formula is in other
format, it can be brought into disjunctive normal form in a
preprocessing step. The conjunctions of literals are called
\emph{clauses} and they correspond to the actions in the \lc
calculus. The prover's state is a \emph{tableau}, which is a graph
whose nodes are labelled with literals. In each step, we select a leaf
node and a clause which contains a literal which is the negation of
the label of the selected node (possibly after unification). The nodes
of the selected clause are added to the tableau graph with edges
pointing to them from the selected node. A constructed tableau tree
proves the input formula if it holds that all paths contain a pair of
literals that are each other's negation.

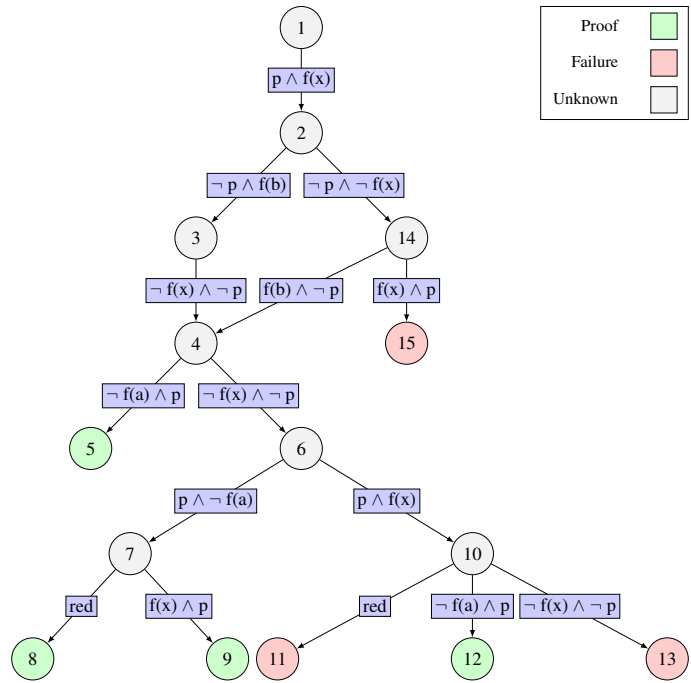
\begin{figure}[htb]
  \centering
  \scalebox{0.7}{
  \begin{tikzpicture}[
    node/.style={circle, draw, fill=gray!10, minimum size=8mm, inner sep=0pt},
    edge label/.style={rectangle, draw, fill=blue!20, inner sep=2pt},
    level 1/.style={sibling distance=2.5cm},
    level 2/.style={sibling distance=4.0cm},
    level 3/.style={sibling distance=2.5cm},
    level 4/.style={sibling distance=4.0cm},
    level 5/.style={sibling distance=6.5cm},
    level 6/.style={sibling distance=3.7cm},
    level distance=2.0cm,
    edge from parent/.style={draw, -latex}
    ]

  \node[node] {1}
    child {node[node] {2}
      child {node[node] {3}
        child {node[node](4) {4}
          child {node[node, fill=green!20](5) {5}
          edge from parent
          node[edge label] {$\neg$ f(a) $\land$ p}}
          child {node[node](6) {6}
            child {node[node](7) {7}
              child {node[node, fill=green!20](8) {8}
              edge from parent
              node[edge label] {red}}
              child {node[node, fill=green!20](9) {9}
              edge from parent
              node[edge label] {f(x) $\land$ p}}
            edge from parent
            node[edge label] {p $\land$ $\neg$ f(a)}}
            child {node[node](10) {10}
              child {node[node, fill=red!20](11) {11}
              edge from parent
              node[edge label] {red}}
              child {node[node, fill=green!20](12) {12}
              edge from parent
              node[edge label] {$\neg$ f(a) $\land$ p}}
              child {node[node, fill=red!20](13) {13}
              edge from parent
              node[edge label] {$\neg$ f(x) $\land$ $\neg$ p}}
            edge from parent
            node[edge label] {p $\land$ f(x)}}
          edge from parent
          node[edge label] {$\neg$ f(x) $\land$ $\neg$ p}}
        edge from parent
        node[edge label] {$\neg$ f(x) $\land$ $\neg$ p}}
      edge from parent
      node[edge label] {$\neg$ p $\land$ f(b)}}
      child {node[node](14) {14}
        child {node[node, fill=red!20](15) {15}
        edge from parent
        node[edge label] {f(x) $\land$ p}}
      edge from parent
      node[edge label] {$\neg$ p $\land$ $\neg$ f(x)}}
    edge from parent
    node[edge label] {p $\land$ f(x)}};
    \draw[-latex] (14) -- node[edge label] {f(b) $\land$ $\neg$ p} (4);

    \begin{scope}[overlay]
      \node[rectangle, draw, fill=white, align=left, anchor=north east] at (current bounding box.north east) {
        \begin{tikzpicture}[
          legend node/.style={rectangle, draw, minimum size=5mm, inner sep=2pt, font=\small},
          legend text/.style={font=\small, align=left}
          ]

          \node[legend node, fill=green!20] (n1) at (1.0,0) {};
          \node[legend text] (t1) at (0,0) {Proof};
  
          \node[legend node, fill=red!20] (n2) at (1.0,-0.7) {};
          \node[legend text] (t2) at (0,-0.7) {Failure};
  
          \node[legend node, fill=gray!10] (n3) at (1.0,-1.4) {};
          \node[legend text] (t3) at (0,-1.4) {Unknown};
        \end{tikzpicture}
      };
    \end{scope}
  \end{tikzpicture}
  }
  \caption{Complete \lc search DAG of a simple theorem
    (pelletier21). Each node represents a \lc tableau, which is a
    directed tree whose nodes are labelled by first order
    literals. Tableau trees are provided in Figure~\ref{fig:tableaux_appendix7}.}
  \label{fig:search_dag_appendix}
\end{figure}

For convenience we repeat the search DAG from
Figure~\ref{fig:search_dag} in Figure~\ref{fig:search_dag_appendix}.

\begin{figure}[ht]
  \centering
  \scalebox{0.7}{
  \begin{tikzpicture}[
    node/.style={circle, draw, fill=gray!10, minimum size=9mm, inner sep=0pt},
    edge label/.style={rectangle, draw, fill=blue!20, inner sep=2pt},
    sibling distance=2.5cm,
    level distance=2.0cm,
    edge from parent/.style={draw, -latex}
    ]

    \node at (0,1.5) {\textbf{State 1}};

  \node[node] {$\neg$\#};
  \end{tikzpicture}
  \hspace{2cm}
  \begin{tikzpicture}[
    node/.style={circle, draw, fill=gray!10, minimum size=9mm, inner sep=0pt},
    edge label/.style={rectangle, draw, fill=blue!20, inner sep=2pt},
    sibling distance=2.5cm,
    level distance=2.0cm,
    edge from parent/.style={draw, -latex}
    ]

    \node at (0,1.5) {\textbf{State 2}};

  \node[node] {$\neg$\#}
    child {node[node] {\#}
    edge from parent}
    child {node[node] {p}
    edge from parent}
    child {node[node] {f(x)}
    edge from parent};
  \end{tikzpicture}
  \hspace{2cm}
  \begin{tikzpicture}[
    node/.style={circle, draw, fill=gray!10, minimum size=9mm, inner sep=0pt},
    edge label/.style={rectangle, draw, fill=blue!20, inner sep=2pt},
    sibling distance=2.5cm,
    level distance=2.0cm,
    edge from parent/.style={draw, -latex}
    ]

    \node at (0,1.5) {\textbf{State 3}};
    
  \node[node] {$\neg$\#}
    child {node[node] {\#}
    edge from parent}
    child {node[node] {p}
      child {node[node] {$\neg$p}
      edge from parent}
      child {node[node] {f(b)}
    edge from parent}
    edge from parent}
    child {node[node] {f(x)}
    edge from parent};
  \end{tikzpicture}
  }
  \label{fig:tableaux_appendix1}
\end{figure}

\begin{figure}[ht]
  \centering
  \scalebox{0.7}{
  \begin{tikzpicture}[
    node/.style={circle, draw, fill=gray!10, minimum size=9mm, inner sep=0pt},
    edge label/.style={rectangle, draw, fill=blue!20, inner sep=2pt},
    sibling distance=2.5cm,
    level distance=2.0cm,
    edge from parent/.style={draw, -latex}
    ]

    \node at (0,1.5) {\textbf{State 4}};

  \node[node] {$\neg$\#}
    child {node[node] {\#}
    edge from parent}
    child {node[node] {p}
      child {node[node] {$\neg$p}
      edge from parent}
      child {node[node] {f(b)}
        child {node[node] {$\neg$f(b)}
        edge from parent}
        child {node[node] {$\neg$p}
        edge from parent}
      edge from parent}
    edge from parent}
    child {node[node] {f(x)}
    edge from parent};
  \end{tikzpicture}

  \hspace{2cm}
  \begin{tikzpicture}[
    node/.style={circle, draw, fill=gray!10, minimum size=9mm, inner sep=0pt},
    edge label/.style={rectangle, draw, fill=blue!20, inner sep=2pt},
    level 1/.style={sibling distance=3.0cm},
    level 2/.style={sibling distance=2.0cm},
    level 3/.style={sibling distance=2.5cm},
    level 4/.style={sibling distance=4.0cm},
    level 5/.style={sibling distance=6.5cm},
    level 6/.style={sibling distance=3.7cm},
    level distance=2.0cm,
    edge from parent/.style={draw, -latex}
    ]

    \node at (0,1.5) {\textbf{State 5}};

  \node[node] {$\neg$\#}
    child {node[node] {\#}
    edge from parent}
    child {node[node] {p}
      child {node[node] {$\neg$p}
      edge from parent}
      child {node[node] {f(b)}
        child {node[node] {$\neg$f(b)}
        edge from parent}
        child {node[node] {$\neg$p}
        edge from parent}
      edge from parent}
    edge from parent}
    child {node[node] {f(a)}
      child {node[node] {$\neg$f(a)}
      edge from parent}
      child {node[node] {p}
      edge from parent}
    edge from parent};
  \end{tikzpicture}
  }
  \label{fig:tableaux_appendix2}
\end{figure}

\begin{figure}[ht]
  \centering
  \scalebox{0.7}{
  \begin{tikzpicture}[
    node/.style={circle, draw, fill=gray!10, minimum size=9mm, inner sep=0pt},
    edge label/.style={rectangle, draw, fill=blue!20, inner sep=2pt},
    level 1/.style={sibling distance=3.0cm},
    level 2/.style={sibling distance=2.0cm},
    level 3/.style={sibling distance=2.5cm},
    level 4/.style={sibling distance=4.0cm},
    level 5/.style={sibling distance=6.5cm},
    level 6/.style={sibling distance=3.7cm},
    level distance=2.0cm,
    edge from parent/.style={draw, -latex}
    ]

    \node at (0,1.5) {\textbf{State 6}};

  \node[node] {$\neg$\#}
    child {node[node] {\#}
    edge from parent}
    child {node[node] {p}
      child {node[node] {$\neg$p}
      edge from parent}
      child {node[node] {f(b)}
        child {node[node] {$\neg$f(b)}
        edge from parent}
        child {node[node] {$\neg$p}
        edge from parent}
      edge from parent}
    edge from parent}
    child {node[node] {f(x)}
      child {node[node] {$\neg$f(x)}
      edge from parent}
      child {node[node] {$\neg$p}
      edge from parent}
    edge from parent};
  \end{tikzpicture}

  \hspace{2cm}
  \begin{tikzpicture}[
    node/.style={circle, draw, fill=gray!10, minimum size=9mm, inner sep=0pt},
    edge label/.style={rectangle, draw, fill=blue!20, inner sep=2pt},
    level 1/.style={sibling distance=3.0cm},
    level 2/.style={sibling distance=2.0cm},
    level 3/.style={sibling distance=2.0cm},
    level 4/.style={sibling distance=4.0cm},
    level 5/.style={sibling distance=6.5cm},
    level 6/.style={sibling distance=3.7cm},
    level distance=2.0cm,
    edge from parent/.style={draw, -latex}
    ]

    \node at (0,1.5) {\textbf{State 7}};

  \node[node] {$\neg$\#}
    child {node[node] {\#}
    edge from parent}
    child {node[node] {p}
      child {node[node] {$\neg$p}
      edge from parent}
      child {node[node] {f(b)}
        child {node[node] {$\neg$f(b)}
        edge from parent}
        child {node[node] {$\neg$p}
        edge from parent}
      edge from parent}
    edge from parent}
    child {node[node] {f(x)}
      child {node[node] {$\neg$f(x)}
      edge from parent}
      child {node[node] {$\neg$p}
        child {node[node] {p}
        edge from parent}
        child {node[node] {$\neg$f(a)}
      edge from parent}
      edge from parent}
    edge from parent};
  \end{tikzpicture}
  }
  \label{fig:tableaux_appendix3}
\end{figure}

\begin{figure}[ht]
  \centering
  \scalebox{0.7}{
  \begin{tikzpicture}[
    node/.style={circle, draw, fill=gray!10, minimum size=9mm, inner sep=0pt},
    edge label/.style={rectangle, draw, fill=blue!20, inner sep=2pt},
    level 1/.style={sibling distance=3.0cm},
    level 2/.style={sibling distance=2.0cm},
    level 3/.style={sibling distance=2.0cm},
    level 4/.style={sibling distance=4.0cm},
    level 5/.style={sibling distance=6.5cm},
    level 6/.style={sibling distance=3.7cm},
    level distance=2.0cm,
    edge from parent/.style={draw, -latex}
    ]

    \node at (0,1.5) {\textbf{State 8}};

  \node[node] {$\neg$\#}
    child {node[node] {\#}
    edge from parent}
    child {node[node] {p}
      child {node[node] {$\neg$p}
      edge from parent}
      child {node[node] {f(b)}
        child {node[node] {$\neg$f(b)}
        edge from parent}
        child {node[node] {$\neg$p}
        edge from parent}
      edge from parent}
    edge from parent}
    child {node[node] {f(a)}
      child {node[node] {$\neg$f(a)}
      edge from parent}
      child {node[node] {$\neg$p}
        child {node[node] {p}
        edge from parent}
        child {node[node] {$\neg$f(a)}
      edge from parent}
      edge from parent}
    edge from parent};
  \end{tikzpicture}

  \hspace{2cm}
  \begin{tikzpicture}[
    node/.style={circle, draw, fill=gray!10, minimum size=9mm, inner sep=0pt},
    edge label/.style={rectangle, draw, fill=blue!20, inner sep=2pt},
    level 1/.style={sibling distance=3.0cm},
    level 2/.style={sibling distance=2.0cm},
    level 3/.style={sibling distance=2.0cm},
    level 4/.style={sibling distance=2.0cm},
    level 5/.style={sibling distance=6.5cm},
    level 6/.style={sibling distance=3.7cm},
    level distance=1.5cm,
    edge from parent/.style={draw, -latex}
    ]

    \node at (0,1.5) {\textbf{State 9}};

  \node[node] {$\neg$\#}
    child {node[node] {\#}
    edge from parent}
    child {node[node] {p}
    edge from parent}
    child {node[node] {f(x)}
      child {node[node] {$\neg$f(x)}
      edge from parent}
      child {node[node] {$\neg$p}
        child {node[node] {p}
        edge from parent}
        child {node[node] {$\neg$f(a)}
          child {node[node] {f(a)}
          edge from parent}
          child {node[node] {p}
          edge from parent}
        edge from parent}
      edge from parent}
    edge from parent};
  \end{tikzpicture}
  }
  \label{fig:tableaux_appendix4}
\end{figure}

\begin{figure}[ht]
  \centering
  \scalebox{0.7}{
  \begin{tikzpicture}[
    node/.style={circle, draw, fill=gray!10, minimum size=9mm, inner sep=0pt},
    edge label/.style={rectangle, draw, fill=blue!20, inner sep=2pt},
    level 1/.style={sibling distance=3.0cm},
    level 2/.style={sibling distance=2.0cm},
    level 3/.style={sibling distance=2.0cm},
    level 4/.style={sibling distance=4.0cm},
    level 5/.style={sibling distance=6.5cm},
    level 6/.style={sibling distance=3.7cm},
    level distance=2.0cm,
    edge from parent/.style={draw, -latex}
    ]

    \node at (0,1.5) {\textbf{State 10}};

  \node[node] {$\neg$\#}
    child {node[node] {\#}
    edge from parent}
    child {node[node] {p}
      child {node[node] {$\neg$p}
      edge from parent}
      child {node[node] {f(b)}
        child {node[node] {$\neg$f(b)}
        edge from parent}
        child {node[node] {$\neg$p}
        edge from parent}
      edge from parent}
    edge from parent}
    child {node[node] {f(x)}
      child {node[node] {$\neg$f(x)}
      edge from parent}
      child {node[node] {$\neg$p}
        child {node[node] {p}
        edge from parent}
        child {node[node] {f(y)}
      edge from parent}
      edge from parent}
    edge from parent};
  \end{tikzpicture}

  \hspace{2cm}
  \begin{tikzpicture}[
    node/.style={circle, draw, fill=gray!10, minimum size=9mm, inner sep=0pt},
    edge label/.style={rectangle, draw, fill=blue!20, inner sep=2pt},
    sibling distance=2.5cm,
    level distance=2.0cm,
    edge from parent/.style={draw, -latex}
    ]

    \node at (0,1.5) {\textbf{State 11}};

  \node[node] {$\bot$};
  \end{tikzpicture}
  }
  \label{fig:tableaux_appendix5}
\end{figure}

\begin{figure}[ht]
  \centering
  \scalebox{0.7}{
  \begin{tikzpicture}[
    node/.style={circle, draw, fill=gray!10, minimum size=9mm, inner sep=0pt},
    edge label/.style={rectangle, draw, fill=blue!20, inner sep=2pt},
    level 1/.style={sibling distance=3.0cm},
    level 2/.style={sibling distance=2.0cm},
    level 3/.style={sibling distance=2.0cm},
    level 4/.style={sibling distance=2.0cm},
    level 5/.style={sibling distance=6.5cm},
    level 6/.style={sibling distance=3.7cm},
    level distance=2.0cm,
    edge from parent/.style={draw, -latex}
    ]

    \node at (0,1.5) {\textbf{State 12}};

  \node[node] {$\neg$\#}
    child {node[node] {\#}
    edge from parent}
    child {node[node] {p}
      child {node[node] {$\neg$p}
      edge from parent}
      child {node[node] {f(b)}
        child {node[node] {$\neg$f(b)}
        edge from parent}
        child {node[node] {$\neg$p}
        edge from parent}
      edge from parent}
    edge from parent}
    child {node[node] {f(x)}
      child {node[node] {$\neg$f(x)}
      edge from parent}
      child {node[node] {$\neg$p}
        child {node[node] {p}
        edge from parent}
        child {node[node] {f(a)}
          child {node[node] {$\neg$f(a)}
          edge from parent}
          child {node[node] {p}
        edge from parent}
        edge from parent}
      edge from parent}
    edge from parent};
  \end{tikzpicture}

  \hspace{2cm}
  \begin{tikzpicture}[
    node/.style={circle, draw, fill=gray!10, minimum size=9mm, inner sep=0pt},
    edge label/.style={rectangle, draw, fill=blue!20, inner sep=2pt},
    sibling distance=2.5cm,
    level distance=2.0cm,
    edge from parent/.style={draw, -latex}
    ]

    \node at (0,1.5) {\textbf{State 13}};

  \node[node] {$\bot$};
  \end{tikzpicture}
  }
  \label{fig:tableaux_appendix6}
\end{figure}

\begin{figure}[ht]
  \centering
  \scalebox{0.7}{
  \begin{tikzpicture}[
    node/.style={circle, draw, fill=gray!10, minimum size=9mm, inner sep=0pt},
    edge label/.style={rectangle, draw, fill=blue!20, inner sep=2pt},
    level 1/.style={sibling distance=3.0cm},
    level 2/.style={sibling distance=2.0cm},
    level 3/.style={sibling distance=2.0cm},
    level 4/.style={sibling distance=2.0cm},
    level 5/.style={sibling distance=6.5cm},
    level 6/.style={sibling distance=3.7cm},
    level distance=2.0cm,
    edge from parent/.style={draw, -latex}
    ]

    \node at (0,1.5) {\textbf{State 14}};

  \node[node] {$\neg$\#}
    child {node[node] {\#}
    edge from parent}
    child {node[node] {p}
      child {node[node] {$\neg$p}
      edge from parent}
      child {node[node] {$\neg$f(y)}
      edge from parent}
    edge from parent}
    child {node[node] {f(x)}
    edge from parent};
  \end{tikzpicture}

  \hspace{2cm}
  \begin{tikzpicture}[
    node/.style={circle, draw, fill=gray!10, minimum size=9mm, inner sep=0pt},
    edge label/.style={rectangle, draw, fill=blue!20, inner sep=2pt},
    sibling distance=2.5cm,
    level distance=2.0cm,
    edge from parent/.style={draw, -latex}
    ]

    \node at (0,1.5) {\textbf{State 15}};

  \node[node] {$\bot$};
  \end{tikzpicture}
  }
  \caption{Tableau trees}
  \label{fig:tableaux_appendix7}
\end{figure}

\end{document}